# Digital Imaging Mass Spectrometry

Running title: Digital Imaging Mass Spectrometry

Casimir Bamberger[1], Uwe Renz[2], and Andreas Bamberger[2]

[1] Department of Chemical Physiology, The Scripps Research Institute, 10550 North Torrey Pines, La Jolla, CA 92037, USA.

[2] Physics Institute, Albert-Ludwigs University of Freiburg, Hermann-Herder-Strasse 3a, 79104 Freiburg, Germany.

***Address reprint requests to:***

Dr. Casimir Bamberger

Department of Chemical Physiology, The Scripps Research Institutes,

10550 North Torrey Pines, La Jolla, CA 92037, USA

Phone: +1-858-784-7645

Fax: +1-858-784-8883

cbamberg@scripps.edu

## Abstract

Methods to visualize the two-dimensional distribution of molecules by mass spectrometric imaging evolve rapidly and yield novel applications in biology, medicine, and material surface sciences. Most mass spectrometric imagers acquire high mass resolution spectra spot-by-spot and thereby scan the object's surface. Thus, imaging is slow and image reconstruction remains cumbersome. Here we describe an imaging mass spectrometer that exploits the true imaging capabilities by ion optical means for the time of flight mass separation. The mass spectrometer is equipped with the ASIC Timepix chip as an array detector to acquire the position, mass, and intensity of ions that are imaged by MALDI directly from the target sample onto the detector. This imaging mass spectrometer has a spatial resolving power at the specimen of $(84\pm35)$ μm with a mass resolution of 45 and locates atoms or organic compounds on a surface area up to ~2 cm$^2$. Extended laser spots of ~5 mm$^2$ on structured specimens allowed parallel imaging of selected masses. The digital imaging mass spectrometer proves high hit-multiplicity, straightforward image reconstruction and its potential for high-speed readout at 4 kHz or more. This device demonstrates a simple way of true image acquisition like a digital photographic camera. The technology may enable a fast analysis of biomolecular samples in near future.

## Introduction

Imaging mass spectrometry is a widely used tool for detecting the molecular composition of sample surfaces [1-6] with a two-dimensional resolution in the µm range [7]. The technology has been proven valuable for many fields of application, ranging from diagnostics in microchip production to analysis of tissue samples. Mass spectrometric images visualize the spatial distribution of molecules with a set of four parameters: two for the spatial dimensions, one for molecules' molecular mass, and one for its abundance.

The commonly used imaging methodology is scanning or micro probing, which first records mass and intensity of molecules or atoms spot-by-spot, and then reconstructs their two-dimensional distribution in a mass spectrometric image [8]. Here, an alternative approach is presented: The two-dimensional distribution of molecules or atoms is acquired in the first step, and their abundance is accumulated by subsequent measurement cycles. The advantage is the parallel acquisition of mass spectrometric data from an extended sample surface [9]. A separation of masses with the time of flight (TOF) principle favors this purpose, since the spatial information can be mapped onto the detector through ion optics. For example, this approach is realized in mass spectrometers of the TRIFT series (Physical Electronics) in the stigmatic mode: Blankers in the ion path set a narrow m/z-range and the spatial distribution of ions is derived from a read out of a phosphor screen by a CCD camera [10, 11]. Alternatively, delay-line read out may be utilized to record the time of flight of spatially separated ions in a single measurement cycle [12]. However, array detectors of high multiplicity offer an attractive and straight-forward realization.

Indeed, the recent development of a time sensitive, highly pixilated detector (Timepix), which was first conceived for 3D-reconstruction of particle tracks [13, 14] and is meanwhile used for photon detection [15, 16], provides a perfect solution. Herein we describe, for the first time to our knowledge, the application of the Timepix detector for imaging mass spectrometry. Recently, the precursor to the Timepix, the Medipix was tested for its application to mass spectrometric imaging, although lacking temporal resolving capabilities [17]. The Timepix detector covers a sensitive surface of $14 \times 14$ mm$^2$ with a matrix of $256 \times 256$ pixels. Each pixel occupies $55 \times 55$ μm$^2$ and detects either charge through time over threshold (TOT) or the time of arrival (TIME) of a signal.

For a proof of principle a linear time of flight mass spectrometer was assembled. Its basic elements were a MALDI driven ion source, a drift space with a focusing ion optics, a multi channel plate (MCP), and the Timepix. The electron signal generated by ions at the MCP covers several pixels of the Timepix, which allows for neighboring pixels to record TOT and TIME in an alternating fashion. As a benefit the simultaneous recording of the electron signal with several pixels allows the reconstruction of the time of arrival with higher precision [18].

The following set of experiments shows the feasibility of this digital mass imaging approach. The mass resolution and imaging properties of the Timepix are presented and the results are discussed within the scope of possible improvements for biological and biomedical applications.

## Material and Methods

### *The Time of Flight Mass Spectrometer and one Event Cycle*

The imaging mass spectrometer apparatus comprises a $N_2$ laser with optics, sample plate with MALDI analyte, drift volume, and detection system as depicted in Figure 1a and described in supplemental materials and methods. The recipient, the acceleration stage (73 mm), the drift space (117 mm) and the MCP, which is part of the detection system, have been retrieved and modified from an experimental setup used previously to analyze the dynamics of molecular interactions [19].

The laser pulse set the reference time and triggered the shutter gate for the Timepix chip (see supplemental material and methods for detailed description of the Timepix). The logic for each pixel provided optionally two modes of operation TIME and TOT (Figure 1b). A shutter gate of fixed length (~10 µs) was applied at the time the laser had fired accepting a charge signal above threshold. From this moment on the 13.5 bit register stored the number of clock cycles (clock speed between 70 and 80 MHz) until either the end of the shutter gate was reached (TIME) or the charge signal dropped below threshold (TOT).

In the final experiment with the extended laser area the grid terminating the drift space was removed to avoid possible interference of micro-lensing with experiments using a highly structured target, thereby slightly modifying the ion optics (Supplemental materials and methods, and supplemental Figure 1).

### Data Analysis

The events were processed with MATLAB 7 (The Mathworks) separating the two modes TIME and TOT and searching for contiguous areas of hits originating from ion signals, subsequently called clusters (see also supplemental materials and methods). The intensity of the ion signal was determined from the sum of the TOT counts in the cluster and the spatial position from the centre of gravity. All clusters, above a predetermined cluster size, were stored along with its spatial position, signal intensity, and time of arrival data for further analysis.

## Results

In order to show the applicability of the Timepix detector for biomolecular imaging mass spectrometry, the imaging mass spectrometer was equipped with focusing ion optics for imaging purposes. An initial set of mass spectra was acquired without activation of the Einzel lens, i.e. in a non-focusing mode and without the diaphragm. MALDI-matrix embedded Cesium (m/z ~133) or Caffeine (m/z ~194), or a mixture of both was deposited on the target sample plate. Depending on the energy of the UV-laser pulse, multiple Caffeine or Cesium ions were extracted simultaneously by MALDI from the target sample in one event.

### Cluster Shapes and Multiplicity

As expected for MALDI ionization, the number of ions detected varied depending on laser intensity and location on the target sample plate. Events recording up to 200 ions resulted in mostly well-separated clusters generated from individual ions as shown in Figure 2a. It demonstrates the capability of the device to record events with high ion multiplicity, which is limited only by excessive overlap of clusters. The separation of clusters depends on the cluster size and the multiplicity (see supplemental results). Small cluster sizes are preferred in order to achieve high multiplicities resulting from increased laser intensities. Because the events shown are recorded in non-focusing mode of the mass spectrometer, the spatial distribution of clusters reflects the dispersion of ions caused by the initial transverse momentum obtained during the MALDI ionization process.

For each recorded event the number of observed clusters was determined. Here more than five adjacent, recording pixels were required for a cluster. Figure 2b depicts a close-up of the matrix TIME for several clusters recorded. The clusters in the TIME matrix have column-like shape with a flat plateau extending over several pixels due to the simultaneous arrival time of the electrons. The corresponding clusters in the TOT matrix are shaped in Gaussian profiles reflecting the two-dimensional projection of the cloud of electrons leaving the MPC (Figure 2c). The event of Figure 2a is displayed in a three-dimensional representation in Figure 2d, which visualizes that the Caffeine signal is well separated from background.

### *Acquisition of Mass Spectra*

The m/z value for each detected ion was calculated from the corrected TIME signal ("time walk" correction, see supplemental results and supplemental Figure 3a) and all m/z values were plotted in a histogram to obtain the final mass spectrum for a series of events recorded (Figure 3a). In this example, two significant peaks appear, which correspond to single-charged Caffeine (m/z ~195) and Iron (m/z ~56), which represents an ionization by-product from the stainless steel backing of the sample.

In a TOF mass spectrometer, the resolution and the accuracy of mass determination are dependent on (1) the dimensions of the mass analyzer and (2) the precision of the time to digital converter. In order to assess the influence of these parameters for the small time of flight mass spectrometer presented, the shape and position of the Caffeine peak was determined by fitting a Gaussian to the data points (Figure 3b). The average time of flight for Caffeine was 9.610±0.138 μs and a signal dispersion of $\Gamma$ = 115 ns. Based on the full width at half maximum (FWHM) for Caffeine, a mass resolution of 45 m/$\Delta$m was determined at m/z 195 (Figure 3b). It should be mentioned that the resolution was not limited by the clock speed (see supplemental results) and no delayed extraction has been applied which is well known to increase mass precision. The overall calibration of the mass spectrometer is demonstrated using three different ions, namely Iron, Cesium and Caffeine (supplemental Figure 3b).

### *Spatial Resolution of the Detector System*

As shown in Figure 1a the field-free section for the ion drift is terminated by a grounded grid, followed by the post acceleration stage of −2.5 kV. As a result of the field gradient, micro-lensing occurred at the detector plane causing the Cesium ion intensities to peak in a periodic pattern when observed over a large number of events (Figure 3c). Using this image of the grid structure, it was possible to estimate the spatial resolution of the combined MCP-Timepix detector system. The analysis of the $Cs^+$ ion intensity pattern with properly chosen angle of projection revealed periodic intensity maxima with a step size of 0.22 mm, which closely matched the periodicity of the grid. Based on the lateral discrimination capability determined from the Cesium intensity pattern, the detector provides an estimated spatial resolution of $\sigma = 36$ μm. Indeed, the spatial resolution is smaller than the lateral extension of a single detector pixel (55 μm), which is due to the fact that the center of gravity is determined with cluster sizes being spread over several pixels.

The imaging capability of the detector section only was visualized in an additional experiment: A copper-coated PC board with 0.2 mm pinholes separated by ≥ 0.6 mm depicting the letters of the word "IMAGE" was positioned near the grid at the end of the drift tube. The passage of mass selected Cs ions through the pinholes reveals the word IMAGE on the detector plane with $10^3$ events (Figure 3d).

This first set of experiments showed that a mass spectrometric detector based on the Timepix is capable of spatially resolving a distribution of ions while determining m/z values at the same time. The analysis of individual clusters recorded in large numbers with each event allowed a fast data acquisition and collection of mass spectra. The simultaneous measurement of TIME and TOT can improve the precision in the m/z-measurement introduced by the finite rise time of the input circuit.

### *Imaging of Laser Spots on the Target Surface*

The instrument provided also a two-dimensional resolution at the target side, which is a prerequisite for imaging. As mentioned above an Einzel lens mapped the ion-optical image of the target plate onto the detector surface. A SIMION 7 simulation verified that the ion optical image was de-magnified by a factor of 2.2 as a consequence of the acceleration stage and the Einzel lens being positioned near the front end of the drift volume. A diaphragm (∅ 6 mm) was introduced near the entrance grid of the drift tube to keep astigmatic distortions small. Following activation of the Einzel lens, Caffeine-ions generated with a laser spot of small size were focused by the ion optical system onto the detector system in a point like spot (supplemental Figures 4a, b).

To show the imaging capabilities of the mass spectrometer $Cs^+$ ions were generated by MALDI at two distinct positions on the target plate by displacing the laser spot. First, the optical focusing of the laser was checked for its quality by estimating its visual transverse extension at the target plate. It was found to be less than $0.3 \times 0.3$ mm$^2$. After recording the mass spectrum at the first position, the laser spot was moved by displacing the lens vertically along the y-axis of the target plate by ~1.0 mm. Due to geometric constraints for the mounting of the Timepix chip its axes are rotated by 135º relative to the target reference system. Therefore, the imaged spot was displaced in both axes of the Timepix reference frame, – by 0.3 mm along the x-axis (Figure 4a) and 0.2 mm along the y-axis (Figure 4b). The resulting ion optical image for $Cs^+$-ions was shifted diagonally by about 0.37 mm on the Timepix detector with an average standard deviation of the Gaussian fits for both spots of 38.1 µm. Concluding from the distinct positions of two spots the imaging mass spectrometer can resolve two spatially separated positions at the sample surface.

Subsequently we determined the spatial resolution and its error with respect to the sample plate. To this end, a more refined estimation of the average standard deviation was performed taking into account, that micro-lensing occurs at the detector stage as described above. By considering the maximum contribution of micro-lensing to the change of the spot width, the resolution with a systematic error was determined to (38.1±16.0) µm. Due to the demagnification of the system the resolution at the sample plate achieved in this experiment is (84±35) µm. A possible contribution due to the MPCs' finite channel diameter of <10 µm is neglected. Whereas a spatial resolution of the digital imaging mass spectrometer is demonstrated in this experiment, the precision of the spatial resolution might be affected by additional factors such as the extension and spatial intensity variations of the laser spot on the target sample plate.

### *Imaging of a Two-dimensional Structure on the Target by Extended Illumination*

The following set of experiments addresses the simultaneous imaging of an extended target plate area in macroscopic scale. For an illumination of an extended surface, the Nitromite $N_2$-laser was replaced by a high power $N_2$-laser providing effective beam energy up to 2 mJ. This allowed for an increase of the illuminated area to an approximate rectangle of ~(5 × 2) mm$^2$ (see supplemental Figure 5b).

The first experiment assessed the general imaging capability of the mass spectrometer with the grid at the end of the drift volume in place. The target consisted of Cesium embedded in a MALDI matrix deposited on a circular plateau of 3 mm ∅, which is separated from the remainder of the flat target plate by a circular grove of 1 mm width (Figure 5a, inset). During target preparation the matrix crystallized preferentially close to the rims of the grove. Imaging of the macroscopic structure on the target plate was achieved along the major axis of the defocused UV-laser beam directly, and along the minor axis by step-wise displacement of the beam between each measurement cycle while the number of measured events was kept equal and the ion optics unaltered. All data acquired was merged to one dataset, which finally reproduced the original ring-like deposition of Cs on the sample plate for the $Cs^+$ line in the mass spectrum (Figure 5a). The same experimental procedure revealed an even distribution of Iron background from an ion image taken from the untreated flat stainless steel plate (supplemental Figure 5a). The spatial frequency distribution of $Cs^+$ signals correlates with the density of $Cs^+$/MALDI-matrix as observed by a light image of the target plate surface. Note that the signal of $Cs^+$ shows a rarified intensity in the center of the disk similar to the Cesium deposit seen in the photograph of the target. The correlation observed between the initial deposition of Cesium and the mass spectrometric image indicates that the digital imaging mass spectrometer can be used to collect information about the spatial distribution of molecules or atoms in a quantitative manner.

In a second experiment, the spatial resolution was tested by imaging isolated spots of analyte deposited onto a stainless steel plate. The setup was modified so micro-lensing was suppressed (See supplemental material and methods, and supplemental Figure 1b). A MALDI target with a regular spacing of spots was produced by ink jet printing [20] of Cesium-containing MALDI matrix onto the stainless steel target plate. Dots were deposited in a rectangular grid with a pitch of 1.6 mm onto the Fe backing (Figure 5b, inset). Figure 5b shows the spatial distribution of the Cesium-MALDI-matrix initially deposited by the printer (white spots). The nominal regular structure of Cesium depositions (yellow disks) and the spatial distribution of Cesium signal recorded by the Timepix detector (red dots) is superimposed on the photograph of the sample plate. The laser image was displaced on the target once by ~0.5 mm during image acquisition and the Iron background signal for one of the two laser positions is shown in supplemental Figure 5b. Regions of high Cesium content closely match the original depositions of the Cesium-MALDI-matrix mixture.

Both sets of experiments show the feasibility to image two-dimensional samples with an extended laser spot and the mass spectrometric images reproduce the initial spatial distributions of the analyte on the target sample plate. The beam energy per pulse limited the sampled surface to about 5-10 mm$^2$. Therefore, we employed the scanning method in order to cover still larger surfaces, but ionization sources with higher energy fluxes are readily available from other laser sources e.g. tripled NdYAG–lasers.

## Discussion and Conclusion

In summary, the Timepix chip was successfully operated as an array detector in a small linear TOF mass spectrometer. The mass spectrometer imaged the spatial distribution of atoms and small organic molecules with sufficiently high mass resolution. It demonstrated that the acquisition of mass spectrometric images is possible with ion detection using front-end digitization and highly pixilated sensors. The Timepix offers the capability of high event rates up to 4 kHz [21] and allows very fast data acquisition with an appropriate interface for data transfer. The experimental setup presented here was not optimized for fast data acquisition, since more advanced USB-based readout modules are currently being developed [22].

The scanning method and delay line readout approach realized in conventional mass spectrometric imaging obtains positional information indirectly. In contrast, highly pixilated detectors directly record all four parameters necessary to acquire mass spectrometric images (position, m/z value, and intensity) thereby omitting cumbersome reconstruction means. Key advantages are (1) multi-hit capability up to 200 ions per event with (2) a lateral resolution of 36 μm of the detector stage. We achieved a lateral resolution of 84 μm at the target sample plate, which is mainly constrained by the de-magnifying ion optical system presently used. It should be stressed that the ion optical demagnification in this setup is unfavorable for demonstrating a high resolution power on the sample due to both the extended acceleration stage and as well as the small drift distance. There are ion optical focusing systems with multiple lenses available [23], which would provide higher magnifications thereby increasing the spatial resolution at the target.

Timepix chips can be seamlessly attached at three sides to expand the detector surface (e.g. quadruplets or higher) thereby increasing either the overall multi-hit capability of the system or the spatial resolution at the target depending on the setting of the ion optics. Although the multi-hit capability of a pixel detector clearly offers the advantage of parallel ion detection, it seems to be limited by the need of a finite cluster size in the mixed mode operation to reconstruct an accurate m/z value and position of the ion signal. However, data acquisition in purely TIME sensitive mode offers a spatial reconstruction marginally worse than with a mixed mode operation. In order to get a time walk correction, the number of pixels in TOT-mode may be reduced in favor of an increased time precision. Moreover, variations in cluster size at small MPC signals are not negligible and more sophisticated algorithms for the separation of partially overlapping clusters are developed.

We estimated the speed and limits of data taking based on the multiplicities of registered ions, which might well be in the region of several hundred per laser shot as shown. If only one dominant ion species would be analyzed, the data rate increases approximately 200 fold per laser shot, provided space-like overlap of ion-induced clusters would be acceptable. Taking into account a maximum laser repetition rate of ~4 kHz, the total improvement in data acquisition speed would be a factor 1000 when compared with conventional micro probing. For the Timepix the data volume is 2 Gb/s, which can be handled easily for transfer and storage. The data volume to be processed is sizable, since the parameters of all clusters per readout cycle have to be determined. Typically such a task can be handled by computing farms using optimized algorithms.

In fact, the data volume would be higher when allowing also the detection of ions with different arrival times impinging at the same position of the detector. However, the Timepix is designed in principle with a single hit feature. Of course a Timepix with multi-hit capability for all pixels would eventually avoid the application of narrow mass windows (see also supplemental discussion).

Modern MALDI scanning mass spectrometric imagers achieve a spatial resolution of typically ~150 µm [24] at the sample-stage dictated by the step size. Using a differential scanning method a resolution of less than 10 µm has been demonstrated [7]. With appropriate sample ionization, the digital imaging mass spectrometer presented here can achieve comparable spatial resolution. The spatial resolution at the sample depends on the ion optics of the imaging TOF mass spectrometer. Indeed, time of flight mass spectrometers with a magnification up to 300 are realized in commercially available instruments of the TRIFT-series. The combination of appropriate ion optics with the highly parallel detection principle will promote digital imaging mass spectrometry as high throughput instrumentation in the future.

The digital imaging TOF as presented here is limited in spectral resolution (m/$\Delta$m 45 at 195 m/z). There are two essential reasons: the short drift path (0.12 m) and the lack of velocity focusing. The digital imaging mass spectrometer was not equipped to correct for initial differences in kinetic energy with the delayed extraction principle, which in general improves spectral resolution. In addition, the time to digital converter (TDC) in the Timepix is restricted to ~100 MHz, a factor 100 lower than the time resolution of modern TDCs used in non-imaging TOF mass spectrometers. Therefore, spatial resolution rather than spectral resolution using the Timepix opens an interesting field of mass spectrometric applications.

The MALDI process was adopted to generate ions on an extended surface because MALDI efficiently ionizes biomolecules and is well employed together with the time of flight mass spectrometry for imaging applications [25]. It is not excluded that the MALDI process has some intrinsic limit for the spatial resolution [26] as well as a transformation of the analyte by crystallization seems to be necessary. With respect to imaging applications, the efficiency and reproducibility of the sample ionization is dependent on many variables introduced during sample preparation and ioninzation (see supplemental discussion). Another field of application is SIMS using fullerenes, where an improvement of spatial resolution with ion optical imaging and digital readout with the Timepix is expected.

In conclusion, the digital mass spectrometer presented here proved the principle of highly parallel mass spectrometric image acquisition. Thereby it opens the door for high throughput sample analysis by imaging ions extracted from extended sample surfaces. Based on the achieved spatial ion multiplicity and the potential readout rates the true mass spectrometric imaging principle presented here allows an increase of data taking speed of several orders of magnitude over micro probing. Moreover, the technology may enable a fast analysis of samples spotted in arrays, eventually illuminated with a bundle of laser beams. Although with limited spectral resolution, the digital imaging mass spectrometer may have broad range of applications ranging from biological and medical experimentation, material sciences to forensic analysis.

## Acknowledgements

The provision of Timepix chips through the MEDIPIX-collaboration/EUDET-collaboration and its expertise is greatly appreciated. Many thanks for Andreas Zwerger being helpful for Timepix related chip mounting and software implementation. The workshops of the Physics Institute, University of Freiburg have provided a perfect support. Prof. Bernd v. Issendorff was helpful providing advices and ion optical simulation programs. We are grateful to Prof. Rüdiger Brenn for providing main parts of time of flight spectrometer. The help of Dr. Sandra Pankow throughout all steps of the project is greatly appreciated.

Idea and experimental design were developed by A.B. and C.B., software development and noise-free calibration of pixels was performed by U.R., instrument setup and data taking were performed by A.B. and C.B., manuscript was prepared by A.B. and C.B.

## Legends for Figures

### Figure 1

**Schematic of a time-of-flight mass spectrometer with high multiplicity detector array and representation of one measurement cycle. (a)** The major components of the time-of-flight mass spectrometer are the acceleration stage (horizontal black, thick, long-dashed lines), the drift volume with the Einzel lens between the grids (vertical dotted lines) and the detector section. The black dashed line indicates the path of ions generated off-axis and focused on the Timepix chip. The $N_2$-laser beam (grey arrow) releases ions by MALDI on the sample plate with a focused spot or with an extended area by adjusting the laser optics (Einzel lens). **(b)** The time sequence starts with opening the shutter gate triggered by the laser pulse (vertical arrowhead). TOT determines approximately the charge of the signal and TIME serves to determine the time of flight after correcting for the gate length. The black dashed line indicates the minimal charge necessary to trigger detection whereas black bars represent the opening time of the shutter, the stored number of cycles (TIME or TOT) and deduced time of flight (TOF). Numbers given to the right exemplify the measured value based on the clock cycle schematic visualized at the bottom of the graph.

### Figure 2

**Recordings by the Timepix visualize MALDI events. (a)** The two-dimensional representation of the detector surface shows the TOF matrix (TIME converted to time of arrival) for one event of high multiplicity recorded from a

Caffeine sample with the Einzel lens switched off. Clusters of pixels are visualized on

the detector surface according to a rainbow-colored time code. Red clusters represent

Caffeine and light blue clusters indicate Iron. **(b)** Close up of a selected x-y surface of

the detector for a single laser shot. The time stamp (TIME) visualizes plateau-like

shaped clusters in units of clock cycles. **(c)** The charge distribution (TOT) of the same

close up shows Gaussian shaped clusters. **(d)** Three-dimensional representation of 129

individually reconstructed clusters based on (a) in terms of time walk corrected, color-coded clock cycles.

### *Figure 3*

**Digital imaging mass spectra, mass and spatial resolution, and image

generating capability of the Timepix. (a)** The mass spectrum ($10^3$ events) shows two

peaks (Caffeine and Iron) above background noise. Solid red curves represent

Gaussian fits. **(b)** A Gaussian curve was fitted to the frequency histogram of the drift times measured for Caffeine. The mass resolution measured was m/$\Delta$m = 45 at m/z ~195 (Caffeine+H$^+$). **(c)** The frequency histogram for the Cesium ion signal visualizes the effect of micro-lensing caused by the grid at the end of the drift volume.

The y-coordinate was rotated with respect to the Timepix to match the grid orientation. The peaks are fitted by multiple Gaussians (solid lines) having a resolution of 36 μm. **(d)** The (x,y)-plot displays the lateral distribution of Cesium ions that were acquired with a pinhole mask depicting the word "IMAGE". The mask was positioned at the end of the drift volume.

**Figure 4**

**Two distant spots on the target sample plate are imaged to distinct positions onto the Timepix.** Frequency histograms of Cesium ions show the displacement along the x-axis **(a)** and along the y-axis **(b)** of the Timepix reference coordinate system as the laser focus was shifted by 1.0 mm at the target. Gaussian fits represent the signal intensity (rate of events) before (red) and after (violet) displacement (arrow) of the laser focus.

**Figure 5**

**Mass spectrometric images of a ring-shaped Cesium deposition and single spots containing Cesium were acquired with an extended laser spot. (a)** The mass spectrometric image visualizes the distribution of Cesium deposits on the target plate (inset). The dotted black lines indicate the projection of the target surface area onto the mass spectrometric image. Each color shows one of the 8 consecutive measurements. Between each measurement, the laser spot was shifted vertically with respect to the target plate reference system. **(b)** A Cesium-containing MALDI sample was spotted by an inkjet printer in a regular pattern (white spots) onto a stainless steel sample plate as shown in the photograph (inset). A scaled photographic image of this sample plate was overlaid with (1) the nominal position of sample spots on the sample plate (yellow) and (2) the respective mass spectrometric image of Cesium ions (red) which were acquired from two adjacent positions consecutively illuminated with the laser beam.

# Figure 1

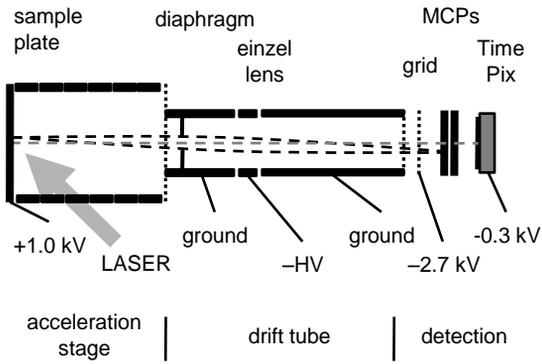

**(a)**

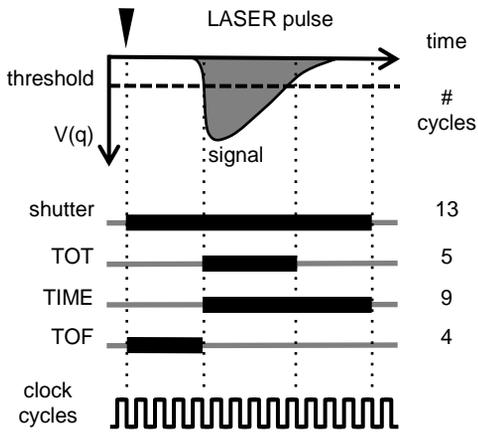

**(b)**

# Figure 2

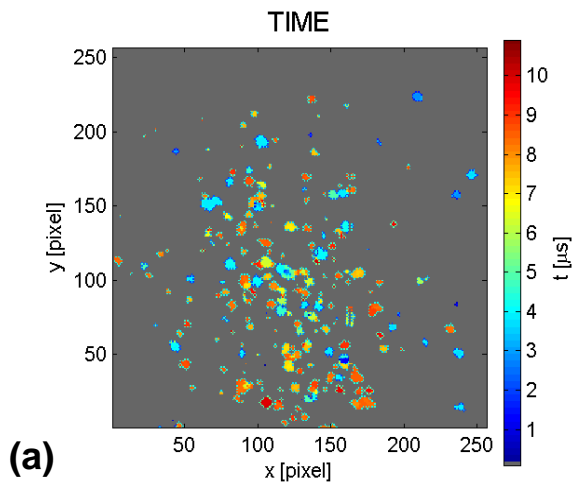

**(a)**

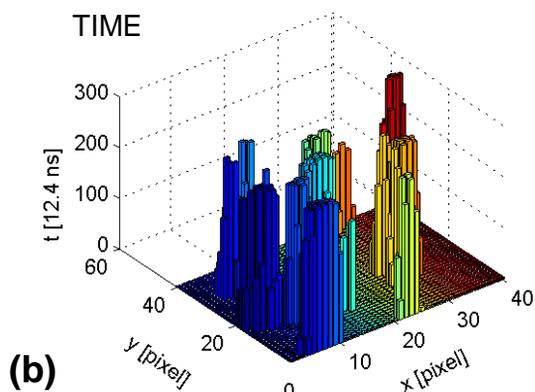

**(b)**

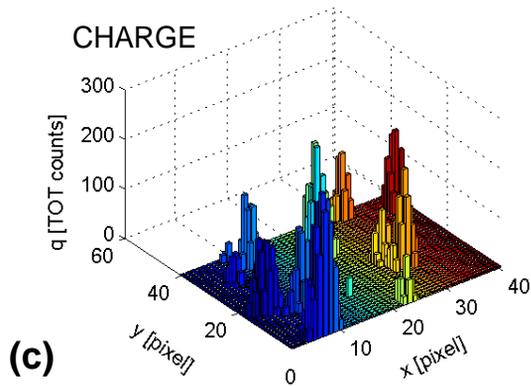

**(c)**

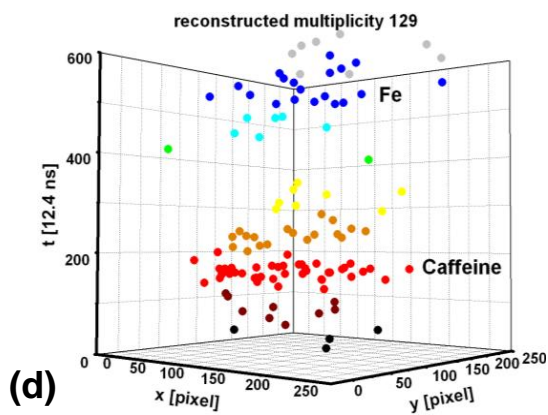

**(d)**

# Figure 3

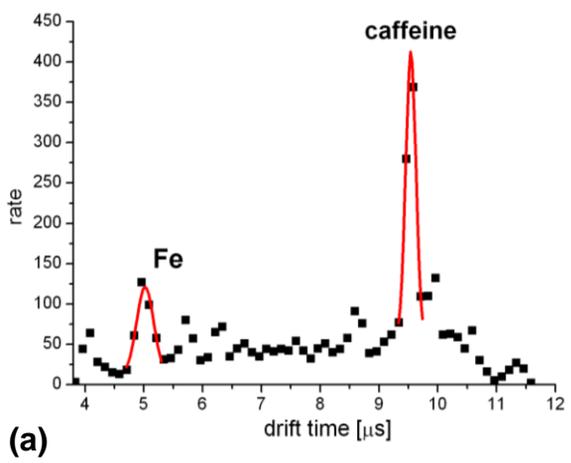

**(a)**

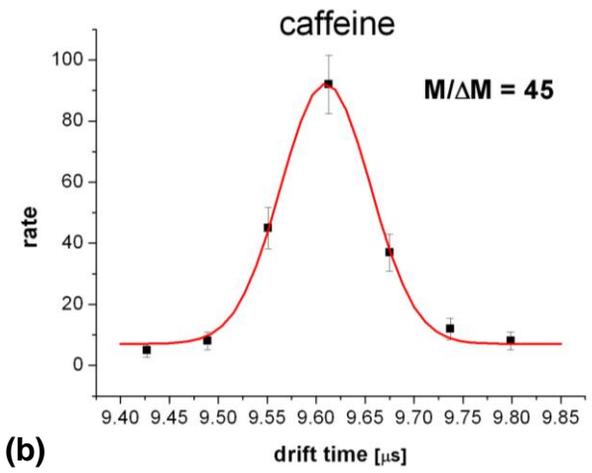

**(b)**

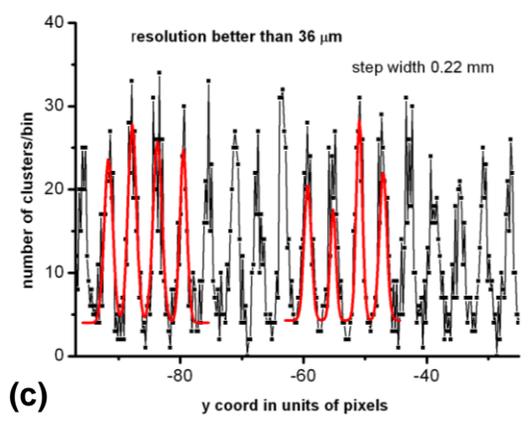

**(c)**

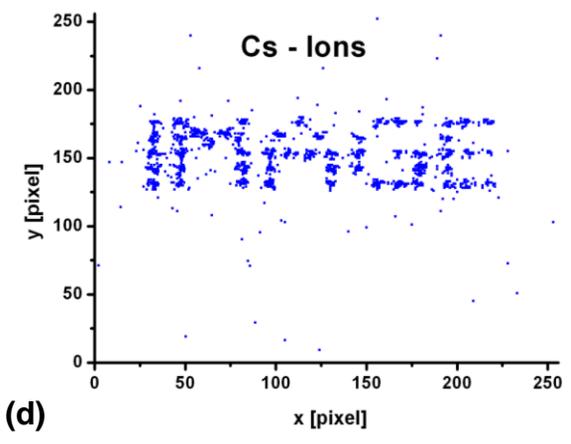

**(d)**

# Figure 4

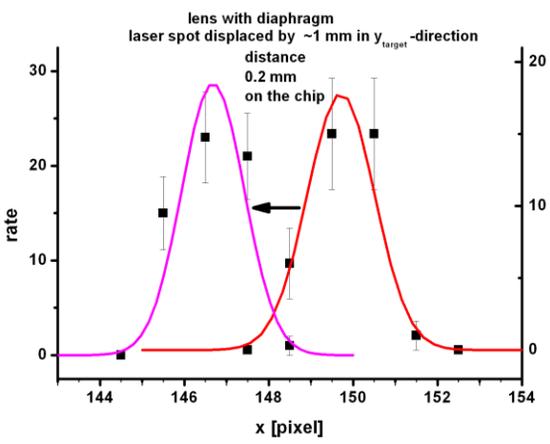

(a)

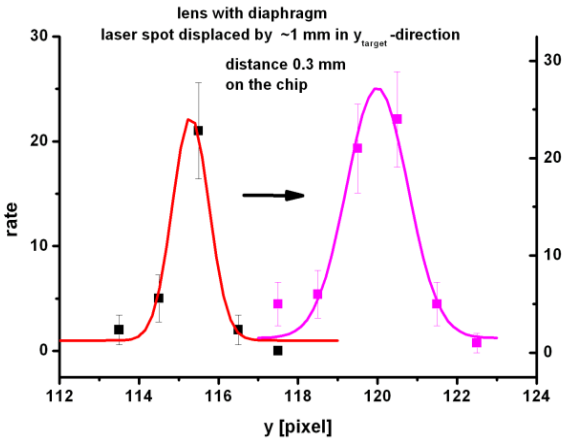

(b)

# Figure 5

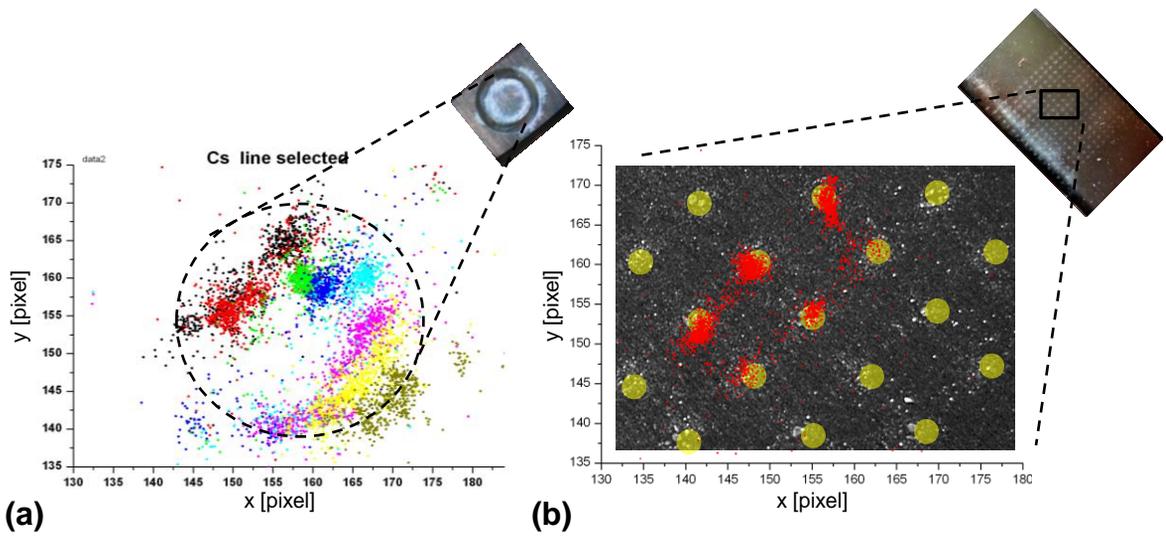

(a)

(b)

# Supplemental Figure 1

**first grid at the end of the drift volume**
**field gradient 0 to 2.5 kV/cm**

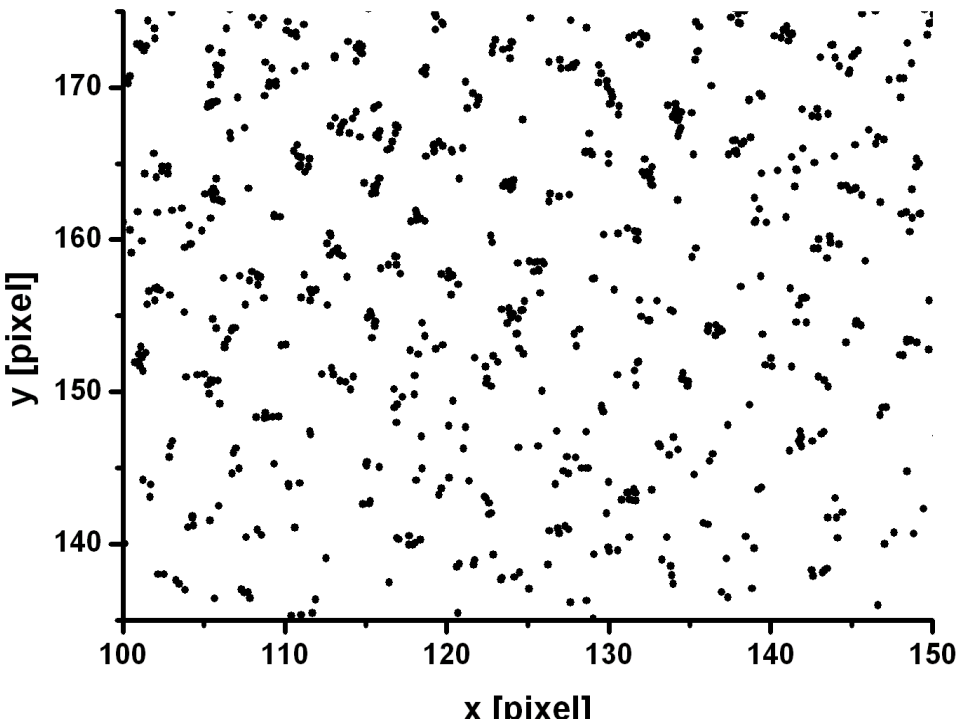

(a)

**no structure without grid**

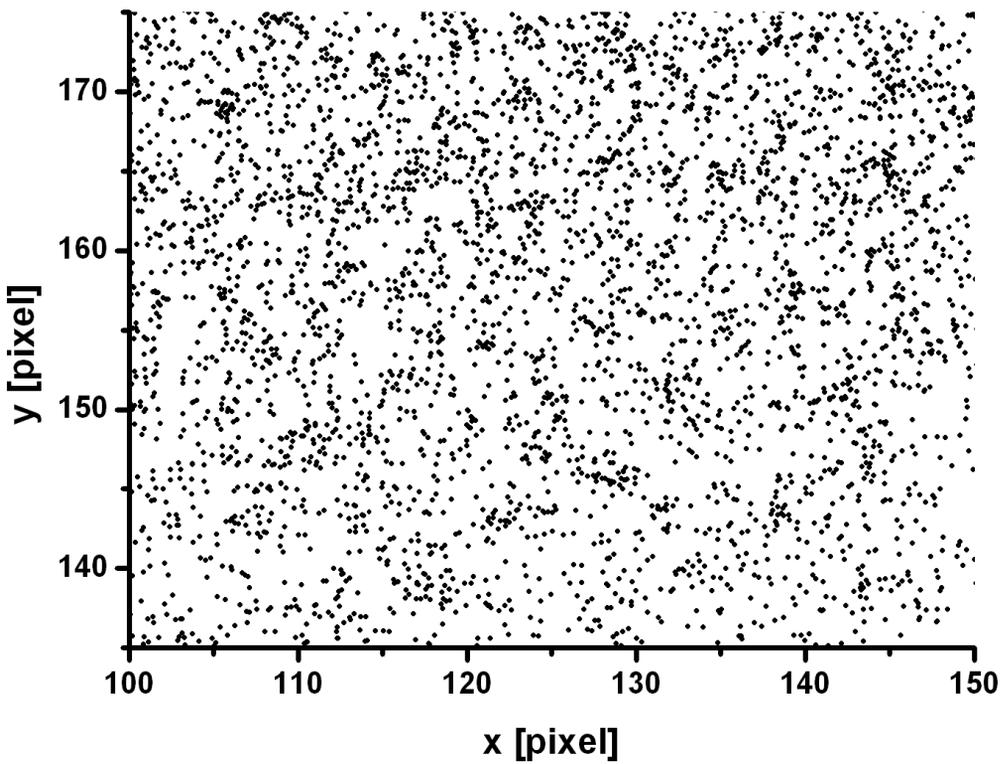

(b)

# Supplemental Figure 2

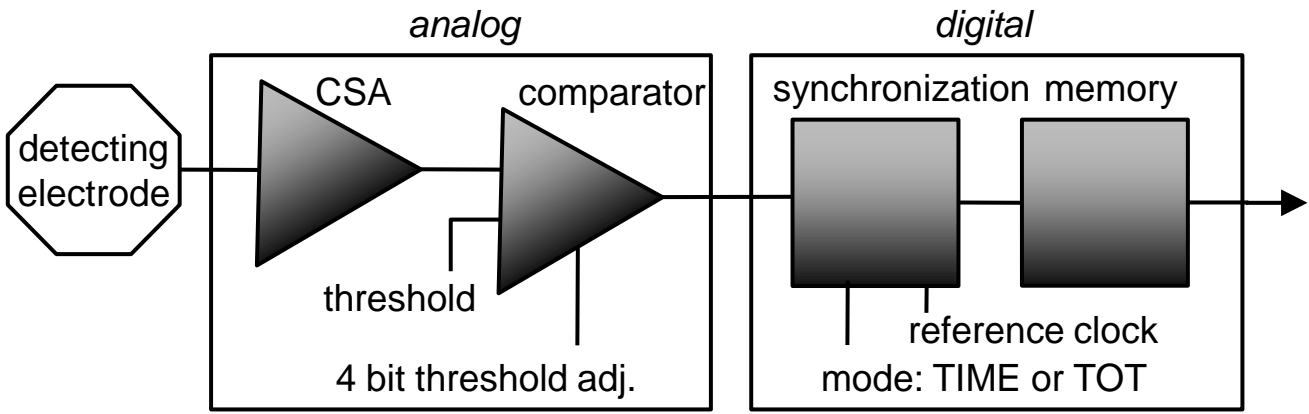

# Supplemental Figure 3

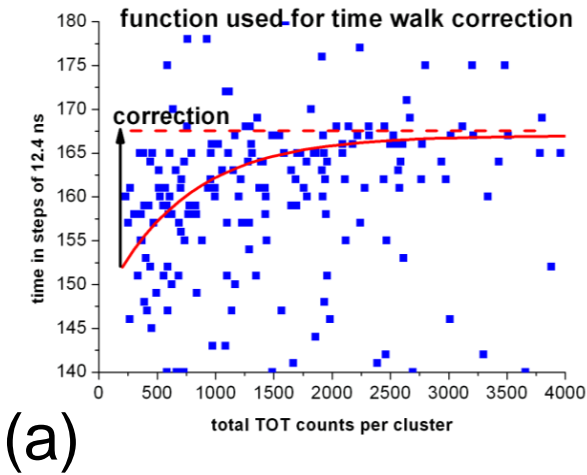

(a)

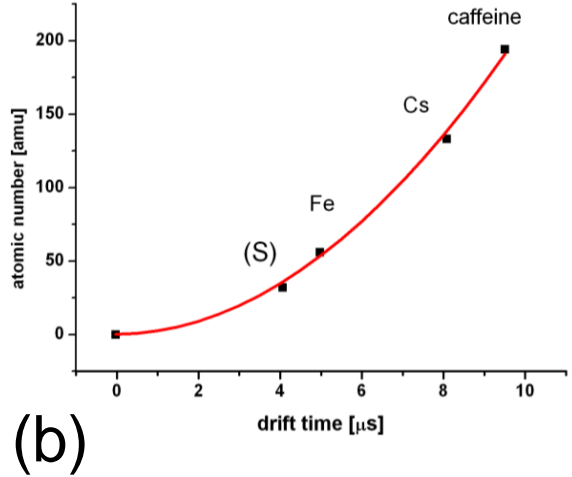

(b)

# Supplemental Figure 4

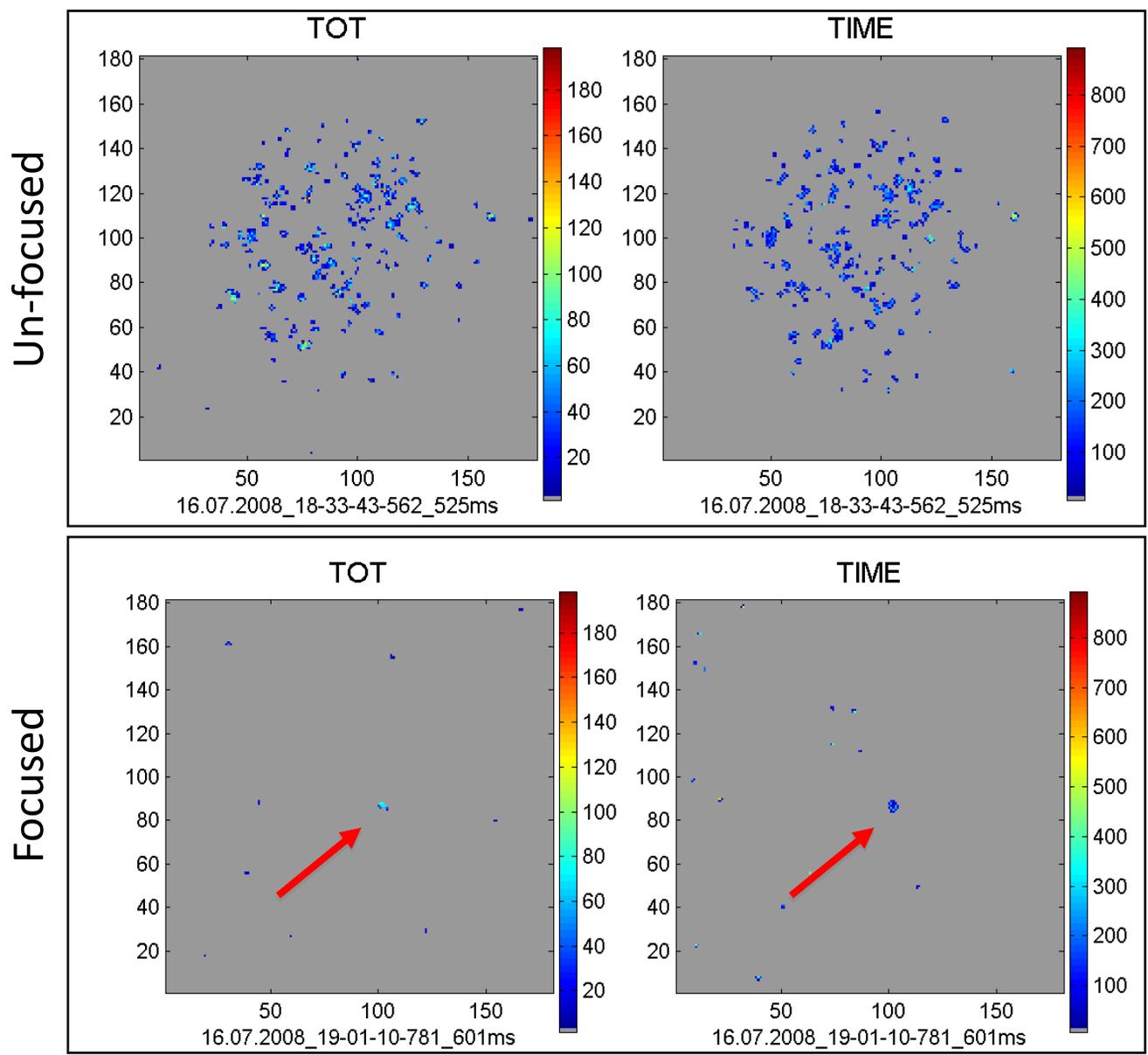

(a)

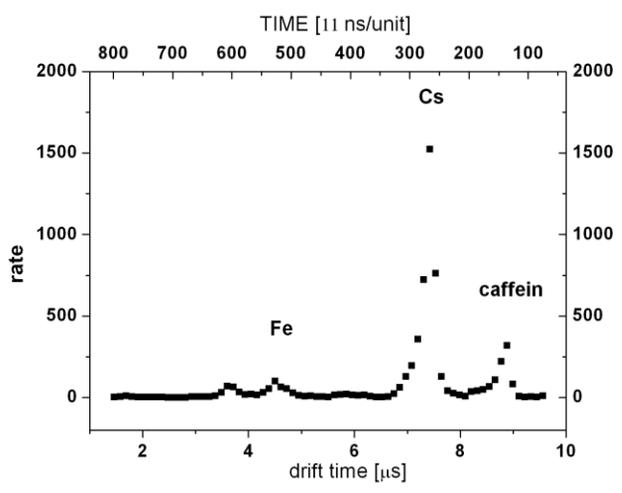

(b)

# Supplemental Figure 5

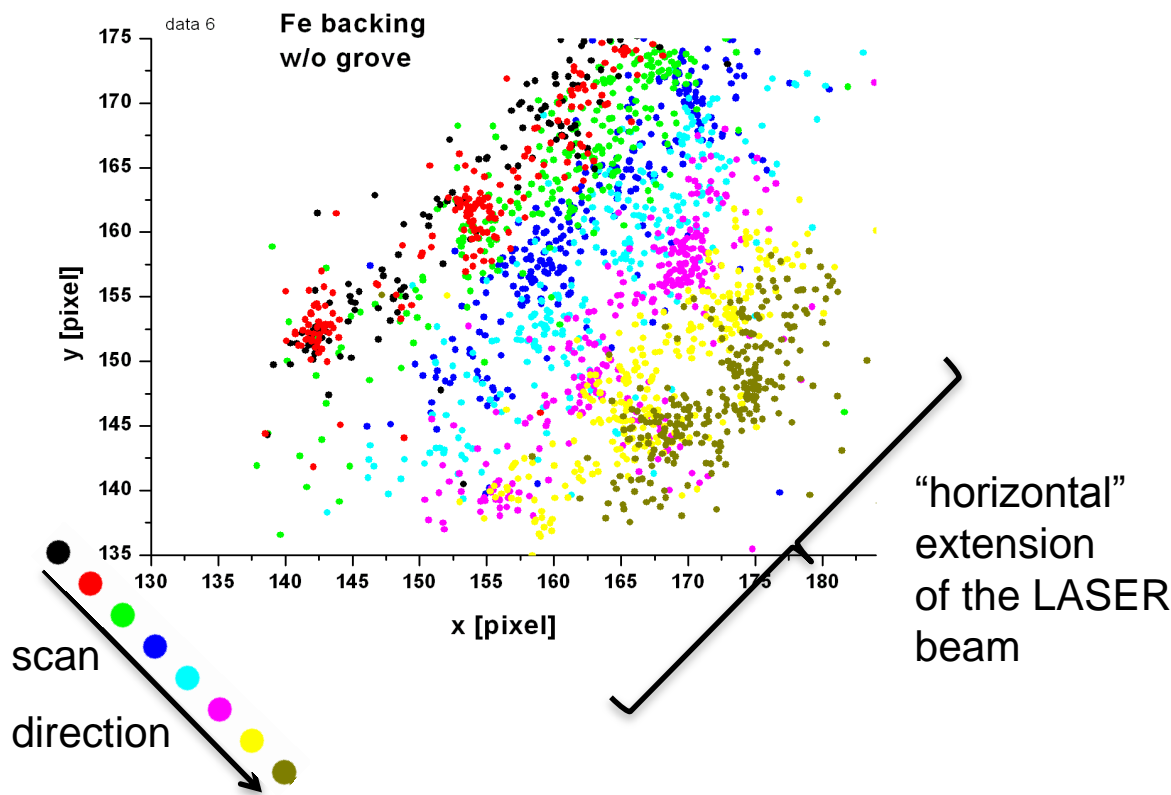

(a)

**Fe backing w/o grove**

"horizontal" extension of the LASER beam

scan direction

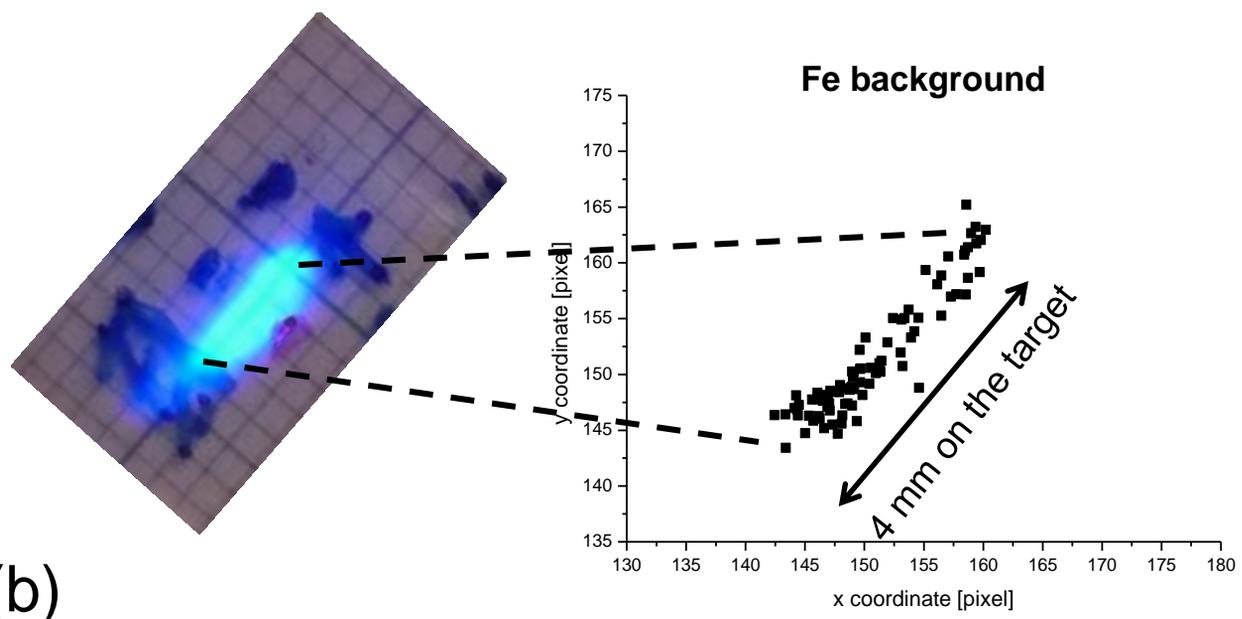

(b)

**Fe background**

4 mm on the target

**Supplemental Information**